\documentclass[10pt, conference]{IEEEtran}
\IEEEoverridecommandlockouts

\usepackage{cite}
\usepackage{amsmath,amssymb,amsfonts}
\usepackage{algorithm}
\usepackage{algpseudocode}
\usepackage{graphicx}
\usepackage[T1]{fontenc}
\usepackage{subfigure}
\usepackage{adjustbox}
\usepackage{textcomp}
\usepackage{xcolor}
\usepackage{verbatim}
\usepackage{amsthm}
\def\BibTeX{{\rm B\kern-.05em{\sc i\kern-.025em b}\kern-.08em
    T\kern-.1667em\lower.7ex\hbox{E}\kern-.125emX}}

\begin{document}
\makeatletter
\newcommand*{\rom}[1]{\expandafter\@slowromancap\romannumeral #1@}
\makeatother

\makeatletter
\setkeys{Gin}{width=\ifdim\Gin@nat@width>\columnwidth
  \linewidth
\else
  \Gin@nat@width
\fi}
\makeatother

\title{Multi-agent Reinforcement Learning for Energy Saving in Multi-Cell Massive MIMO Systems \vspace{-3mm}} 
\author{\IEEEauthorblockN{
Tianzhang Cai\IEEEauthorrefmark{1}, Qichen Wang\IEEEauthorrefmark{1}, Shuai Zhang\IEEEauthorrefmark{1}, \"Ozlem Tu\u{g}fe Demir\IEEEauthorrefmark{2},  and Cicek Cavdar\IEEEauthorrefmark{1}
}
\IEEEauthorblockA{\IEEEauthorrefmark{1}Department of Computer Science, KTH Royal Institute of Technology, Stockholm, Sweden (\{tcai, shuai2, cavdar\}@kth.se)}
\IEEEauthorblockA{\IEEEauthorrefmark{2}Department of Electrical-Electronics Engineering, TOBB ETU, Ankara, Türkiye  (ozlemtugfedemir@etu.edu.tr) \vspace{-4mm}}
}

\vspace{-8mm}

\maketitle
\begin{abstract}
We develop a multi-agent reinforcement learning (MARL) algorithm to minimize the total energy consumption of multiple massive MIMO (multiple-input multiple-output) base stations (BSs) in a multi-cell network while preserving the overall quality-of-service (QoS) by making decisions on the multi-level advanced sleep modes (ASMs) and antenna switching of these BSs. The problem is modeled as a decentralized partially observable Markov decision process (DEC-POMDP) to enable collaboration between individual BSs, which is necessary to tackle inter-cell interference. A multi-agent proximal policy optimization (MAPPO) algorithm is designed to learn a collaborative BS control policy. To enhance its scalability, a modified version called MAPPO-neighbor policy is further proposed.
Simulation results demonstrate that the trained MAPPO agent achieves better performance compared to baseline policies. Specifically, compared to the auto sleep mode 1 (symbol-level sleeping) algorithm, the MAPPO-neighbor policy reduces power consumption by approximately 8.7\% during low-traffic hours and improves energy efficiency by approximately 19\% during high-traffic hours, respectively. 
\end{abstract}


\section{Introduction}


The advent of fifth-generation (5G) cellular networks is set to bring
about a range of enhancements, including lower latency, higher data
rates, and wider connectivity, compared to previous generations. The deployment of a denser network of base stations (BSs) is expected to accommodate
these improvements. However, this densification will result in a considerable increase
in energy consumption, as BSs are the most energy-intensive components
of a wireless network and are responsible for approximately 80\% of
a network's energy consumption \cite{lahdekorpiEnergyEfficiency5G2017}. 
This increase in energy consumption is not sustainable in the long
term, making the deployment of green networks essential. A green network
takes sustainability into account, ensuring that the network
equipment and architecture are energy efficient across varying network
conditions. While 5G presents challenges in terms of energy consumption,
it also offers opportunities to implement new methods for energy conservation \cite{olsson5GrEEnGreen5G2013,peesapatiAnalyticalEnergyPerformance2021}.

In 5G networks, energy efficiency is improved by several techniques such as the separation of data and control plane \cite{olsson5GrEEnGreen5G2013}, traffic adaptive configuration of antenna elements in massive MIMO (mMIMO) BSs \cite{Peesapati2021Q-learningBR, Hossain2018EnergySG}, and management of advanced sleep modes \cite{Masoudi2022DigitalTA}. In \cite{Peesapati2021Q-learningBR}, centralized single agent reinforcement learning is used to configure the bandwidth and number of antenna elements in 5G networks given a static network load. In \cite{Hossain2018EnergySG}, only the number of antenna elements is configured to save energy in a multi-cell mMIMO system with given traffic load. In \cite{Masoudi2022DigitalTA}, reinforcement learning is used to choose the best sleep mode among several options with different sleep durations in a single BS to find a compromise between network performance and energy savings with dynamic traffic arrivals where the future user traffic is unknown. \cite{Masoudi2022DigitalTA} focuses on a single BS and does not consider the impact of inter-cell interference.

Jointly deciding on BS sleep mode management and configuration of antenna elements in a multi-cell environment with dynamically changing traffic remains an open problem and challenge. A multi-cell network requires the collaboration of the BSs to save energy jointly, which may lead to a different cooperative policy than the optimal one for each BS considered separately. Therefore, the required BS control policy must be dynamic, adaptive, and cooperative in nature. In order to tackle the joint sleep mode management and antenna configuration for energy-efficient multi-cell networks, we develop a deep reinforcement learning algorithm to achieve dynamic and cooperative multi-agent BS control. The main contributions of this paper are outlined as follows:
 \vspace{-1mm}
\begin{itemize}
\item We develop a 5G network simulation environment by
mimicking the mobile traffic pattern based on deep packet inspection (DPI) data collected from one of the Swedish network operators.
\item A multi-agent proximal policy optimization (MAPPO) algorithm is designed for multi-agent BS control of antenna switching and sleep mode management in the simulated network in order to cooperatively save network power consumption (PC) without compromising quality-of-service (QoS). 
\item It has been shown that the proposed algorithm consumes less energy compared to the baseline algorithms, including a vanilla system without energy-saving mechanisms and a system with a simple policy that automatically puts idle BSs into the shallowest sleep mode.
\end{itemize}

The rest of the paper is organized as follows.
We present the system model in Section II and elaborate on the proposed MAPPO-based algorithm in Section III. In Section IV, simulations are conducted to demonstrate the superiority of the proposed algorithm in comparison to the two baseline algorithms. Finally, Section V concludes this paper.

\section{System Model}

We consider a multi-cell massive MIMO network with $C$ BSs, where each BS\textbf{ $b_{c}$} for $c=1,\ldots,C$ has the maximum number $M_{\rm max}$ of antennas. At a given time instance, suppose that BS $b_c$ is serving $K_{c}$ user equipments (UEs) connected to its cell with its $M_c\leq M_{\rm max}$ active antennas. Thus there are a total number of $K=\sum_{c=1}^{C}K_{c}$ UEs in this network. Let $c_{k}$
denote the index of the BS serving UE $u_{k}$ for $k=1,\ldots,K$.
We assume that the communication bandwidth is $B$\,Hz for all the BSs. Assuming perfect channel state information is available and zero-forcing precoding is used at each BS by nulling the intra-cell interference, an achievable data rate of UE $u_k$ in a massive MIMO system is
given by  
 \vspace{-2mm}
\begin{equation}
r_{k}=B\log_{2}\left(1+\text{SINR}_{k}\right), \label{eq:rate}
\end{equation}
where the $\text{SINR}_{k}$ is the effective \emph{signal-to-interference-plus-noise ratio} of UE $u_k$, which is given as \cite{marzettaFundamentalsMassiveMIMO2016}
\begin{equation}
{\rm SINR}_{k}=\frac{\mathcal{S}_{k}}{\mathcal{I}_{k}+\sigma^2}=\frac{(M_{c_{k}}-K_{c_{k}})\beta_{c_{k},k}p_{c_{k},k}}{\sum\limits _{\substack{c\neq c_{k}}
}\beta_{c,k}p_{c}+\sigma^2}, \label{eq:sinr}
\end{equation}
where $\mathcal{S}_{k}$, $\mathcal{I}_{k}$, and $\sigma^2$ are the desired signal
power, interference power, and noise power, respectively. The $\beta_{c,k}\geq 0$ is the large-scale fading coefficient for the channel from BS $b_c$ to UE $u_k$, which is modeled as
\begin{align}
10\log_{10}\beta_{c,k}=\mathrm{PL}\left(d_{c,k}\right)+\chi_{c,k},
\end{align}
where $\mathrm{PL}(\cdot)$ models the distance-dependent path loss in decibel scale, $d_{c,k}$ is the distance between
BS $b_{c}$ and UE $u_{k}$, and $\chi_{c,k}$ is a random variable modeling
log-normal shadow fading that follows a Gaussian distribution $\mathcal{N}\left(0,\sigma^2_{\text{SF}}\right)$. The
$p_{c,k}$ is the transmit power allocated to UE $u_{k}$ by BS $b_{c}$
($p_{c,k}=0$ for $c\ne c_{k}$) and $p_{c}=\underset{k: c_{k}=c}{\sum}p_{c,k}$
is the total output power of BS $b_{c}$. Assuming that the maximum allowable transmit power per BS is $p_{\rm max}$, then $p_{c}\leq p_{\rm max}, \forall c$.
The noise power is given by $\sigma^2=B\cdot10^{\left(N_{0}+N_{B}\right)/10}$,
where $N_{0}$ is the noise power spectral density, and $N_{B}$ is
a constant called ``noise figure'' depending on the type of BS hardware, both in decibel scale. 

\subsection{Traffic Model}

We process the DPI data of one mobile operator to generate a realistic simulation environment. The DPI engine records the traffic flows
in a time granularity of two seconds and differentiates between about
300 applications, such as HTTPS, Facebook, WhatsApp, Youtube, etc.
This allows us to classify the sites based on their traffic patterns
in different categories according to the delay requirement of the
provided services. According to the 3GPP specification document TS
23.501 \cite{Specification23501}, there is a certain packet delay
budget for each category of network service to specify its QoS requirement, and based on that, we divide the delay budgets into three categories: i) delay stringent category with 50\,ms packet delay budget including real-time gaming and communication; ii) delay sensitive category with 150\,ms packet delay budget including multimedia streaming and social networking; and iii) delay tolerant category with 300\,ms packet delay budget including other web applications, mail, file hosting, and tunnel and remote access services. 

For each service category $z$, we aggregate
the network flows into time slots of 20 minutes, take the weekly average, and then calculate the temporal-spatial traffic density $\kappa_{z,t}$ in each time slot $t$ in $\text{Mb/s/km}^{2}$. Each UE has a certain size of traffic demand and a respective delay budget. The arrival rate of the UEs is defined so as to mimic the traffic pattern in each of the delay categories. We assume each UE only requests a single network flow and we model
the traffic flows as a Poisson process with an average arrival rate
of $\lambda_{z,t}$, approximately the probability of a new UE to be
generated in the area in a timestep. Its value follows the equation
\begin{align}
\lambda_{z,t}=\frac{\kappa_{z,t}A}{x_f}\Delta t,
\end{align}
where $A$ is the simulation area in km$^2$, $x_f$ is the file size requested by each UE in Mb, and $\Delta t$ is the duration of each discretized timestep.

In each timestep, UEs may arrive at any random location in the area. During the period of service, UE $u_k$'s demand size,
denoted by $x_{k}(t)$ in time slot $t$, will decrease with the corresponding data rate $r_{k}(t)$. If UE $u_{k}$
is not being served by any BS, its data rate is $r_{k}(t)=0$. We denote the delay,
i.e., the elapsed time since the arrival of UE $u_{k}$, by $\tau_{k}(t)$ at time slot $t$,
and then the demand size of that UE over time follows the equation 
\begin{align}
x_{k}\left(t\right)=\text{max}\left(0,x_{f}-\sum_{i=1}^{\tau_{k}(t)/\Delta t}r_{k}\left(i\cdot \Delta t\right)\Delta t\right),
\end{align}
where $x_f$ is the initial demand size of each UE. A UE will stay in the environment until its demand has been finished,
i.e., $x_{k}(t)=0$, or its delay has reached the budget. Since each service category $z$ has a different delay budget $\tau_{\text{max}}^{(z)}$,
the maximum time UE $u_{k}$ can stay in the environment is $\tau_{\text{max}}^{(z_k)}$, where $z_k$ is the service category of UE $u_k$. Then UE $u_{k}$ requires a minimum data rate $r_{\text{req},k}={x_{f}}/{\tau_{\text{max}}^{(z_k)}}$ on average.
At the time when UE $u_{k}$ quits the network, we denote its remaining
demand as $\chi_{k}$ and when $\chi_{k}>0$, it will be ``dropped'', which degrades the QoS of UE $u_{k}$. The network performance in
our simulation is mainly based on the \emph{drop ratio}  $\frac{\chi_{k}}{x_{f}}$, which takes values between $0$ and $1$.
\subsection{Advanced Sleep Modes}
The introduction of 5G networks offers increased adaptability and energy efficiency compared to earlier generations. Leveraging the capabilities of 5G technology, BSs can now access deeper sleep levels for optimal power reduction, while still effectively addressing UE requirements and network operation challenges \cite{salemManagementAdvancedSleep2019}. Utilizing the so-called \emph{advanced sleep modes (ASMs)}, 5G BSs can reduce PC significantly without negatively impacting overall system performance or the UE experience. Employing ASMs  involves sending the BS into deeper sleep levels by successively deactivating components with longer activation delays. Consequently, lower overall PC is achieved while increasing the wake-up latency. As documented in \cite{debaillieFlexibleFutureProofPower2015}, there are four sleep modes (SM 1-4) characterized by activation delays,  each having a different PC discount factor. The values of these parameters are adopted from \cite{tombazEnergyPerformance5GNX2016,peesapatiAnalyticalEnergyPerformance2021} and shown in Table
\ref{tab:sms}.

\begin{table}[H]
\caption{Sleep modes. \label{tab:sms}}
\begin{centering}
\begin{tabular}{|l|c|c|c|c|}
\hline 
Sleep level $s$ & 0 & 1 & 2 & 3\tabularnewline
\hline 
\hline 
Activation latency $\Delta_{s}$ & 0\,ms & 1\,ms & 10\,ms & 100\,ms\tabularnewline
\hline 
PC discount factor $\delta_{s}$ & 1 & 0.69 & 0.5 & 0.29\tabularnewline
\hline 
\end{tabular}
\par\end{centering}
\end{table}

\subsection{PC Modeling of Massive MIMO BS with ASM}
Following \cite{hossainEnergyEfficientLoadAdaptiveMassive2015}, we can model the total PC of BS $b_{c}$ as
\begin{align}
P_{c}\left(K_{c},M_{c}, s_c\right)=\delta_{s_{c}}\left(M_{c}P_{{\rm PA}}\left(p_{a}\right)+P_{{\rm BB}}\left(K_{c},M_{c}\right)+P_{{ o}}\right),\nonumber
\end{align}
where $P_{\rm PA}(p_a)$ gives the PC of a power amplifier (PA)
when its average output power is $p_a$, $P_{\mathrm{BB}}\left(K_{c},M_{c}\right)$
is the baseband signal processing power, which is a function of both the number of active antennas and served UEs. $P_{{o}}$
includes the load-independent PC for site cooling, control signal,
DC-DC conversion loss, etc. The sleep level of BS $b_c$ is denoted by $s_c$ and when BS $b_c$ is active, we have $s_c=0$ with $\delta_{s_c}=1$, i.e., the original PC model in \cite{hossainEnergyEfficientLoadAdaptiveMassive2015}. When $s_c>0$,  the transmit power of the BS will be zero and we have $M_c=K_c=0$, but there will be a certain idle power consumption, which is  multiplied by the discount factor $\delta_{s_c}$.

\subsection{Intra-Cell Power Allocation}

After the connection is established, the BS allocates its output
power using a relatively simple method. As a first-order approximation,
other conditions given the same, the data rate of  UE $u_{k}$ satisfies
$r_{k}=O\left(\log_{2}p_{c_{k},k}\right)$, as can be seen in
(\ref{eq:rate})-(\ref{eq:sinr}). Therefore, suppose UE $u_{k}$
requires a minimum data rate of $r_{\text{min},k}=\frac{x_{k}(t)}{\tau_{\text{max}}-\tau_{k}(t)}$ at a particular time slot $t$,
then the power allocated to UE $u_k$ is given as
\begin{align}    p_{c_k,k}=\frac{2^{r_{\text{min},k}}}{\sum\limits_{\substack{j=1,\\c_j=c_k}}^K2^{r_{\text{min},j}}}p_{c_k}.
    \end{align}

\section{MAPPO-based Multi-cell ASM and Antenna Switching Algorithm }

Since there are multiple BS agents in the multi-cell network environment,
and cooperative actions are required to achieve the desired outcome
of balancing between energy efficiency and service quality, the problem
is formulated in a \emph{multi-agent reinforcement learning (MARL)}
framework.
With some specific modifications, the PPO algorithm has demonstrated
strong performance in cooperative multi-agent games \cite{yuSurprisingEffectivenessPPO2021}
like the Starcraft micromanagement challenge (SMAC) or Google Research
Football \cite{kurachGoogleResearchFootball2020}.
Thus, we propose a MAPPO-based algorithm to jointly determine the ASM  and antenna switching policy across multiple cells.

\subsection{Action Space and State Space}

The set of actions a BS $b_c$ can take in the simulation is defined as 
\begin{equation}
\mathcal{A}^c=\mathcal{A}_m\times\mathcal{A}_s,\label{eq:actions}
\end{equation}
where 
\begin{equation}
\mathcal{A}_m=\left\{ -4,0,4\right\} 
\end{equation}
stands for the options of antenna switching: $-4$ corresponds to switching off
4 antennas, 0 indicates no antenna switching, and 4 implies switching on 4 additional antennas.
\begin{align}
   \mathcal{A}_s=\left\{ 0,1,2,3\right\}  
\end{align}
represents actions for switching to the corresponding sleep level
$s\in\mathcal{A}_s$. Since there are $C$ BSs in total,
the number of agents in the network system is $C$ and we denote
the agents as $b_{1},b_{2},\ldots,b_{C}$.
We define the \emph{joint action space} $\mathcal A=\mathcal A^{1}\times \mathcal A^{2}\times\ldots\times \mathcal A^{C}$
as the set of joint actions $\left(a_{1},a_{2},\ldots,a_{C}\right)$.
In our problem, all the BS agents have the same action space $\mathcal{A}$,
as defined in \eqref{eq:actions}, thus $\mathcal A=\mathcal{A}^{c}$.

The state space consists of the total PC, the number of finished/dropped UE requests, the average of (average data rate / required data rate) for finished/dropped UE requests in the last action interval, the required sum rate of all idle/queued/served UEs, current sum rate of all UEs, and individual BS observations regarding PC, the number of active antennas, and sleep level. 

\subsection{DEC-POMDP}
In reality, each BS agent can only observe a part of the whole environment state. Specifically, a BS agent is only aware of
UEs within its own cell coverage. If a certain BS increases its output power
to improve the QoS of its covered UEs, it may generate more interference to UEs outside of its coverage. Thus the best action based on the observation of a single agent is often not the optimal one in terms of the state of the whole environment.

Formally speaking, if the environment is in state $s$, an agent $b_{c}$
only has access to its local observations $o_{c}$, which contains
partial or uncertain information of $s$. The \emph{observation space}
$O_{c}$ of agent $b_{c}$ is the set of observations it can get from
any state. We define the \emph{joint observation space} $O=O_{1}\times O_{2}\times\ldots\times O_{C}$
as the set of joint observations $\left(o_{1},o_{2},\ldots,o_{C}\right)$.

If the environment is partially observable, then the observation space
$O$ and an observation model $Z\left(o\mid s\right)=\mathbb{P}\left[o\mid s\right]$
need to be included in the Markov decision process (MDP) framework to define a \emph{partially
observable MDP (POMDP)}. The framework can be further extended to
a \emph{decentralized POMDP (DEC-POMDP)} when there are $C$ agents
acting in the environment without a centralized controller. To simplify
the problem, we assume the observation model is deterministic, i.e.,
without uncertainty. Consequently $Z$ becomes a direct mapping $\mathcal S\rightarrow O$
such that the joint observations in state $s$ is $\left(o_{1},\ldots,o_{C}\right)=Z\left(s\right)$.

In summary, a DEC-POMDP is defined by a tuple $\left(C,\mathcal S,O,\mathcal{A},P,Z,R,\gamma\right)$,
where $\mathcal S$ is the state space, $O$ and $\mathcal{A}$ are the
the joint observation and action space, respectively,  $P\left(s^{\prime}\mid s,a\right)=\mathbb{P}\left[s_{t+1}=s^{\prime}\mid s_{t}=s,a_{t}=a\right]$
is the transition model, $Z\left(o\mid s\right)=\Pi_{c=1}^{C}Z_{c}\left(o_{c}\mid s\right)$ is
the joint observation model, $R: \mathcal {S\times A\times S}\rightarrow\mathbb{R}^{C}$ is
the reward model, and $\gamma$ is the discount
parameter.

Our MARL problem is an instance of DEC-POMDP with two modifications:
shared reward $R:\mathcal S\rightarrow\mathbb{R}$ and deterministic observation
model $Z:\mathcal S\rightarrow O$.

\subsection{Reward Design}
Since the BSs in our multi-cell network environment can be seen as homogeneous agents, we assume the same reward model $R$ for all agents. This reward model is solely dependent on the new state after a transition takes place, i.e., $R_{t}\left(s,a_{t},s^{\prime}\right)=R\left(s^{\prime}\right)$
for $t=1,2,\ldots$ Therefore after each joint action, all agents
share the same reward.
To minimize the total PC of multiple massive MIMO  BSs  while preserving the UEs' QoS, we design the reward function as
\begin{align}
R & =w_{\text{qos}}\xi-w_{\text{pc}}P_N
\end{align}
with
\begin{align}
\xi & =\frac{1}{K}\sum_{k=1}^{K}\xi_{k}=\frac{1}{K}\sum_{k=1}^{K}\begin{cases}
\rho_{k}-1 & \rho_{k}<1,\\
\phi\left(1-\frac{1}{\rho_{k}}\right) & \rho_{k}\ge 1,
\end{cases}
\end{align}
where $P_N$ is the average PC of
the network. Denoting the delay of UE $u_k$ is $\tau_k$ when it quits the network, $\rho_{k}=\frac{r_{\text{avg},k}}{r_{\text{req},k}}$,
where $r_{\text{avg},k}=\frac{x_{f}-\chi_{k}}{\tau_{k}}$ is the average
data rate of UE $u_{k}$. $\xi_{k}$ stands for the reward for the QoS experienced by
UE $u_{k}$. When UE $u_k$ is dropped, $\tau_k=\tau_{\text{max}}^{(z_k)}$,  $r_{\text{avg},k}<r_{\text{req},k}$ and
$\rho_{k}<1$, so $\xi_{k}=\rho_{k}-1$ is a negative value as penalty,
whose \emph{ magnitude}  in this case equals $\frac{\chi_{k}}{x_{f}}$,
\emph{the ratio of dropped data}. When the request is finished, $r_{\text{avg},k}\ge r_{\text{req},k}$
and $\rho_{k}\ge1$, so $\xi_{k}=\phi\left(1-\frac{1}{\rho_{k}}\right)$
is a non-negative reward. Note that in this case $\frac{1}{\rho_{k}}={\tau_{k}}/{\tau_{\text{max}}^{(z_k)}}$,
which can be interpreted as\emph{ the ratio of delay to its budget}.
When the service completely failed the demand, $\xi_{k}=-1$; when
it finished the total demand ``just in time'', $\xi_{k}=0$, thus
no reward nor penalty; while if its data rate was much higher than
required, there is an upper bound of reward, which is $\phi>0$. Generally speaking, we want to set $\phi$ to a small value because in comparison to those
demands failed to be finished in time and thus dropped, the data rates
of those services finished within the delay budget are much less critical
to the QoS. Finally, $w_{\text{pc}}$ and $w_{\text{qos}}$ are parameters
used to balance between the QoS reward and the PC penalty in the reward.
We can also neglect the data rate of a service as long as it finishes
before the delay budget by setting $\phi=0$, provided that other
conditions (PC, drop rate, etc.) are the same.
In the simulations, the values of   $w_{\text{qos}}$, $w_{\text{pc}}$, and $\phi$  are set as 4, 1, and 0.005, respectively.

\subsection{Learning in DEC-POMDP}

The trajectory is extended to include the observations: $\tau=\left(s_{0},o_{0},a_{0},r_{0},\ldots,s_{T-1},o_{T-1},a_{T-1},r_{T-1},s_{T}\right)$,
where $o_{t}=\left(o_{t,1},\ldots,o_{t,C}\right)$ is the joint observations
and $a_{t}=\left(a_{t,1},\ldots,a_{t,C}\right)$ is the joint actions.
Since the agents are homogeneous, we use a single policy network $\pi_{\theta}$
for all agents. This can reduce the \emph{non-stationarity} of the
environment since the policies of other agents as part of the environment
have lower variance. Hence $a_{t,c}=\pi_{\theta}\left(o_{t,c}\right)$.
We also denote $a_{t}=\pi_{\theta}\left(o_{t}\right)$ for convenience. Assuming shared policy and shared reward, the DEC-POMDP is in effect reduced to a POMDP, whose action space is all agents' joint action
space and observation space is their joint observation space.

In MARL, each agent considers the other agents as part of the environment. To optimize its policy, an agent needs to acquire knowledge about the environment, which includes the actions and behaviors of other agents. However, since all agents are simultaneously updating their policies, the dynamics of the environment for a single agent also change. This dynamic nature of the environment can negatively impact or even prevent the convergence of training, which is known as the ``non-stationarity'' problem in MARL.

The \emph{centralized training and decentralized execution (CTDE)}
approach is an effective strategy for addressing the non-stationarity problem in MARL. In this approach, an actor-critic architecture is utilized, where the training phase is centralized, and the execution phase is decentralized. During centralized training, the critic network has access to both the global state of the environment and the local observations of all agents. This centralized perspective allows the critic to evaluate the states under the agents' policies consistently, providing stable evaluations even in the presence of changing dynamics.
Once the training phase is completed, the critic is no longer required, and each agent can independently execute its learned policy in a fully decentralized manner. 
\begin{figure}[t!]
\begin{centering}
\includegraphics[width=0.5\columnwidth]{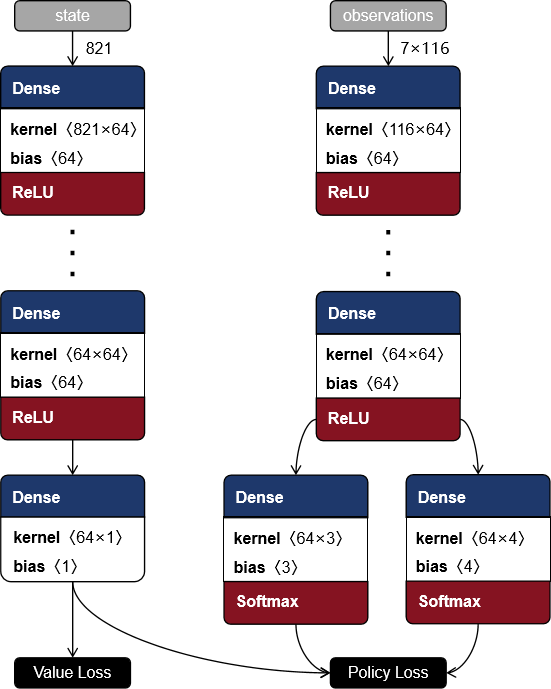}
\par\end{centering}
\caption{Network structure of MAPPO agent.\label{fig:net-structure}}
\end{figure}
\section{Results and Analysis \label{ch:resultsAndAnalysis}}
We consider a square area of $1\text{km}\times1\text{km}$, where $C=7$ BSs $b_{c}$ ($c=0,1,\ldots,6$)
are placed in this area -- one ($b_{0}$) locates in the center
and the other six surround it as a hexagon with a side length of $d_{\text{BS}}=400$\,m. The simulation proceeds
by time-stepping with a step length of $\Delta t=1$\,ms. According to the micro urban model with non-line-of-sight propagation in the 3GPP TR
38.901 report \cite{3GPP_38_901}, the path loss is modeled as
\begin{align}
\mathrm{PL}\left(d_{c,k}\right)=-35.3\log_{10}(d_{c,k})-22.4-21.3\log_{10}(f_{c})
\end{align}
where $f_c=5$\,GHz is the carrier frequency. The standard deviation for the shadow fading is $\sigma_{\text{SF}}=7.82$. Other simulation parameters are shown in Table \ref{tab:net-params}.

\begin{table}[t!]
\caption{Parameters of the Network. \label{tab:net-params}}
\begin{centering}
\begin{tabular}{lc}
\hline 
Parameter & Value\tabularnewline
\hline 
BS antenna height: $h_{\text{BS}}$ & 30\,m\tabularnewline
UE average height: $h_{\text{UE}}$ & 1.5\,m\tabularnewline
Episode time length & 1008\,s\tabularnewline
Delay budgets: $\tau_{\text{max}}^{(1/2/3)}$ & 50/150/300\,ms\tabularnewline
File size: $x_{{f}}$ & 3\,Mbits\tabularnewline
Bandwidth: $B$ & $20$\,MHz\tabularnewline
Noise power spectral density: $N_{0}$ & $-204$\,dB$\cdot$J\tabularnewline
Noise figure: $N_{B}$ & 7\,dB\tabularnewline
Average transmit power per antenna: $p_a$ & 0.1\,W\tabularnewline
Minimum number of antennas: $M_{\text{min}}$ & 16\tabularnewline
Maximum number of antennas: $M_{\text{max}}$ & 64\tabularnewline
Load-independent power: $P_{{o}}$ & 18\,W\tabularnewline
\hline 
\end{tabular}
\par\end{centering}
\end{table}

\begin{figure*}[t]
\begin{minipage}{0.32\textwidth}
    \centering
    \includegraphics{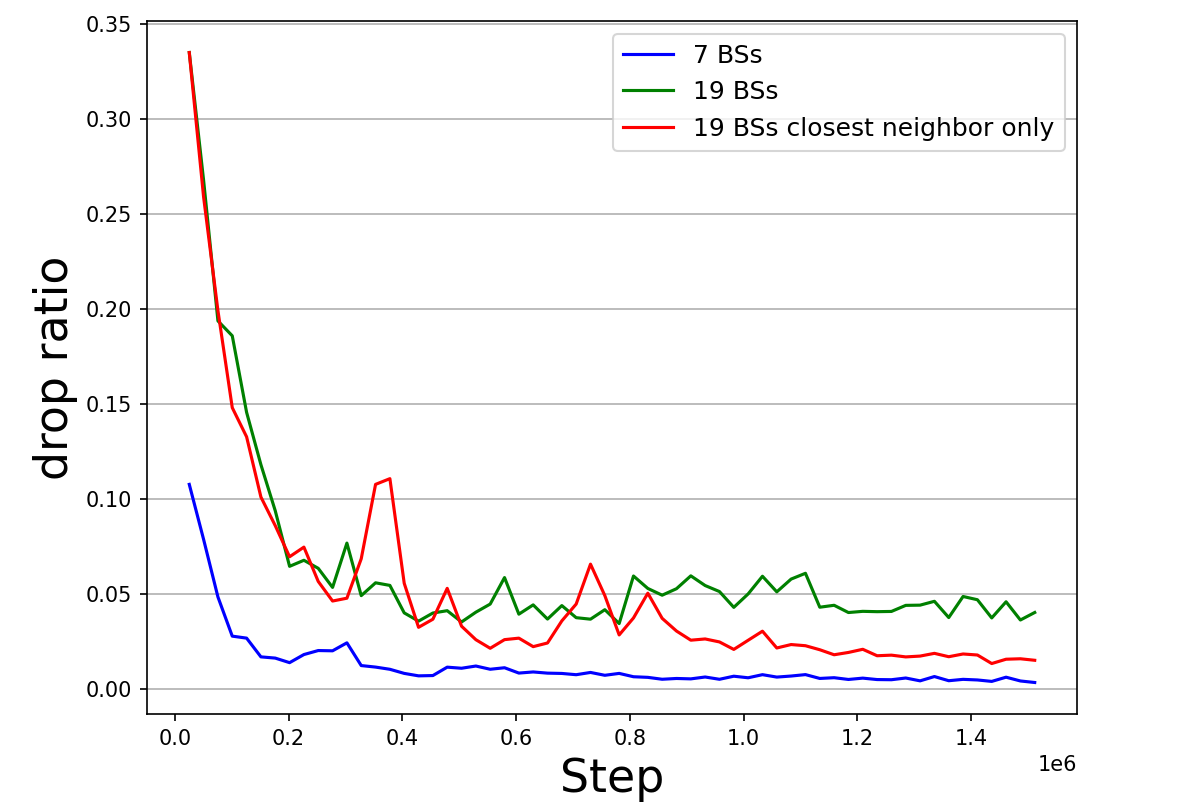}
    \vspace{-8mm}
    \caption{Training drop ratio.  \label{fig:drop-ratio-train}}
\end{minipage}
\begin{minipage}{0.32\textwidth}
    \centering
    \includegraphics{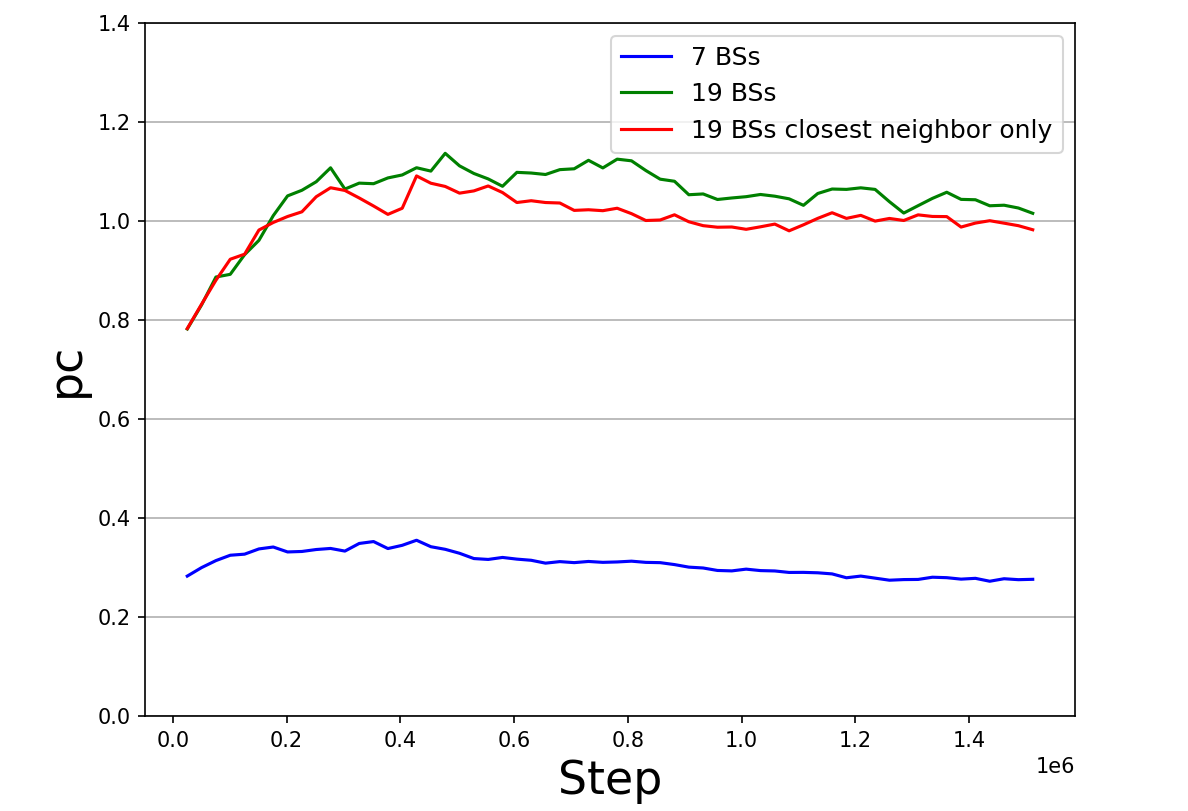}
    \vspace{-8mm}
    \caption{Training power consumption. \label{fig:pc-train}}
\end{minipage}
\begin{minipage}{0.32\textwidth}
    \centering
    \includegraphics{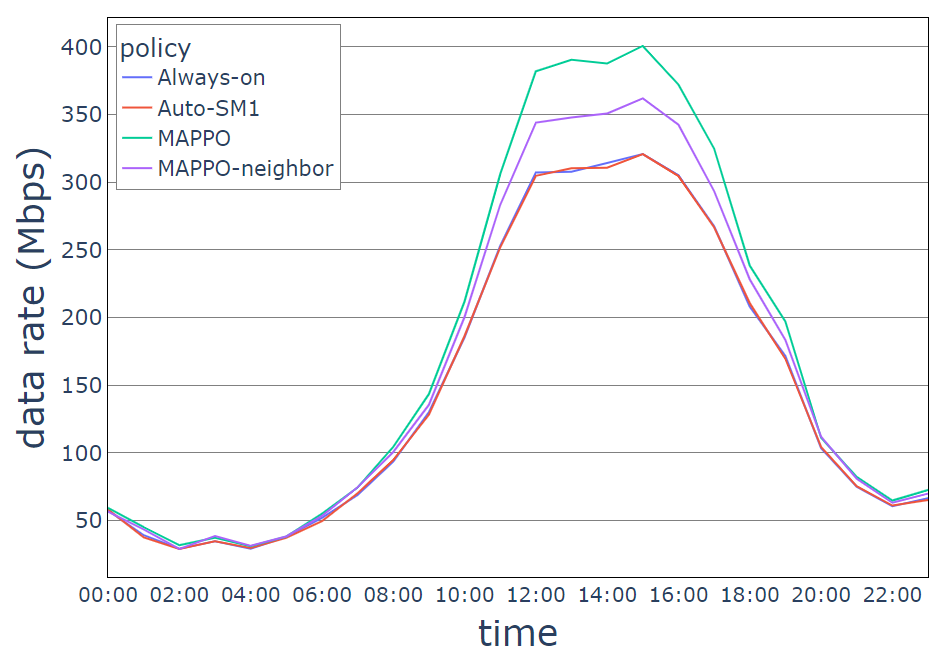}
    \vspace{-8mm}
    \caption{Comparison of sum data rate. \label{fig:data-rate}}
\end{minipage}
\end{figure*}

The batch size $T$, i.e., the length of trajectories used for MAPPO
learning, can have a flexible range of values. However, a larger batch size allows the algorithm to learn more efficiently and make better use of the collected experience \cite{yuSurprisingEffectivenessPPO2021}.
To enable the network to learn from week-long trajectories, we set the batch size equal to the episode length. 
After gathering each trajectory, it is utilized to train
the network for 10 passes (epochs). To avoid the risk of overfitting and maintain stability, rather than dividing the trajectory into mini-batches, the training is performed using full-batch gradient descent. The entire training process concludes after 100 episodes of simulation. Fig.~\ref{fig:net-structure} illustrates the deep neural network structure of the MAPPO agent. Other parameters of the learning network are given in Table \ref{tab:MAPPO-params}.

\begin{table}[t!]
\caption{Parameters of MAPPO learning.} \label{tab:MAPPO-params}
\begin{centering}
\begin{tabular}{lc}
\hline 
Parameter & Value\tabularnewline
\hline 
Agent timestep & 20\,ms\tabularnewline
Batch size $T$ & 50400 steps = 1 episode\tabularnewline
Training episodes $N$ & 100\tabularnewline
Epochs per episode $M$ & 10\tabularnewline
Discount factor $\gamma$ & 0.99\tabularnewline
Actor learning rate $\eta_{\pi}$ & 6e-4\tabularnewline
Critic learning rate $\eta_{v}$ & 5e-4\tabularnewline
Number of mini-batches & 1\tabularnewline
Clip parameter $\varepsilon$ & 0.2\tabularnewline
Generalized
advantage estimation parameter $\lambda$ & 0.95\tabularnewline
Entropy coefficient $c_{e}$ & 0.01\tabularnewline
Huber loss parameter $\epsilon$ & 10\tabularnewline
\hline 
\end{tabular}
\par\end{centering}
\end{table}

To demonstrate the scalability of MAPPO, we also consider a larger area, i.e., doubling the side length of square area and add more BSs outside the 7 BSs with the same hexagonal structure where the total number of BSs reaches 19. Since the PC increases with 19 BSs, the $w_{\text{pc}}$ is decreased to 0.4 to balance the increased PC. Note that the combination of all agents' observations at the critic network results in too much repeated storage of the network state. In order to make model focus more on key features and enhance training stability, we use \emph{closest neighbor only} policy here, which simplifies the size of observations where each BS can only access the information of its 6 closest neighbor BSs, and this modification can reduce the training time to converge by approximately 44.2\%. For the simplification of terminology, we will use \emph{MAPPO-neighbor} to represent \emph{closest neighbor only} in this paper.

We compare the training performance of MAPPO-based algorithm in 7 BSs, 19 BSs, and 19 BSs with \emph{MAPPO-neighbor} policy, respectively. The results are shown in Figs. 2-3.
In Fig. \ref{fig:drop-ratio-train}, the drop ratio of 7 BSs converges to the lowest value due to the simplest environment. Compared with 19 BSs with redundant information, 19 BSs with \emph{MAPPO-neighbor} policy converges to a better value with less fluctuations. 
In Fig. \ref{fig:pc-train}, 7 BSs consume less power than 19 BSs. Besides, we can observe that with \emph{MAPPO-neighbor} policy, the PC of 19 BSs converges to a lower level because of a better training with less repeated input. 

\begin{figure*}[t]
\centering
\begin{minipage}{0.32\textwidth}
    \centering    
    \includegraphics{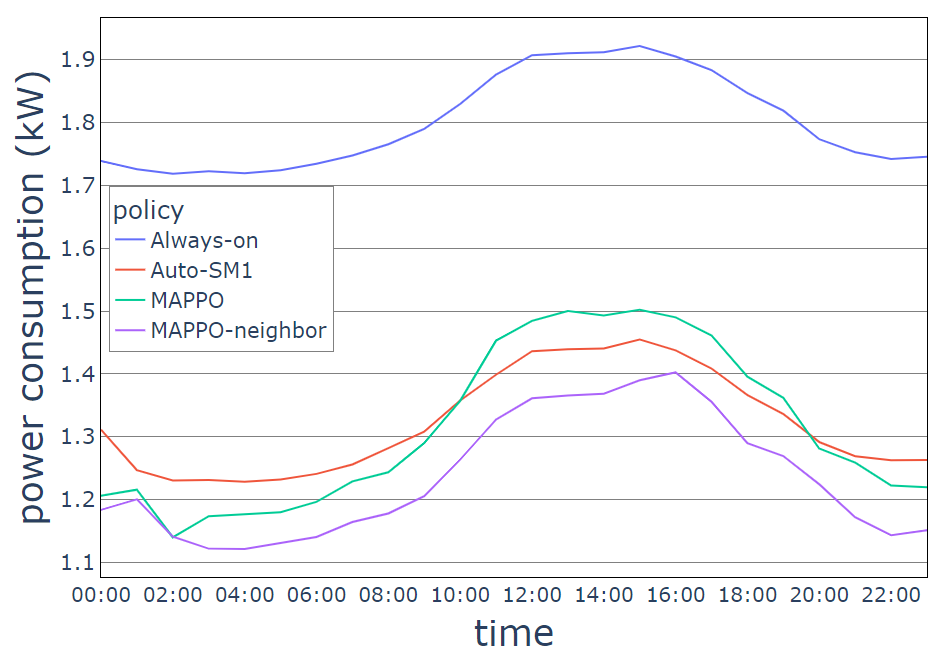}
    \vspace{-8mm}
    \caption{Comparison of power consumption. \label{fig:power}}
\end{minipage}
\begin{minipage}{0.32\textwidth}
    \centering
    \includegraphics{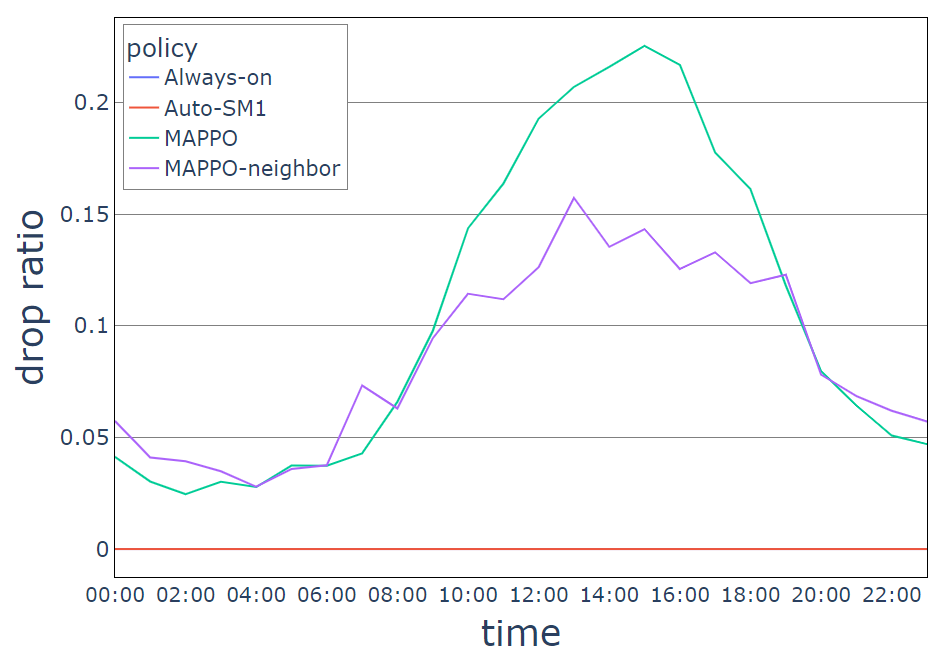}
    \vspace{-8mm}
    \caption{Comparison of drop ratio. \label{fig:drop-ratio}}
\end{minipage}
\begin{minipage}{0.32\textwidth}
    \centering
    \includegraphics{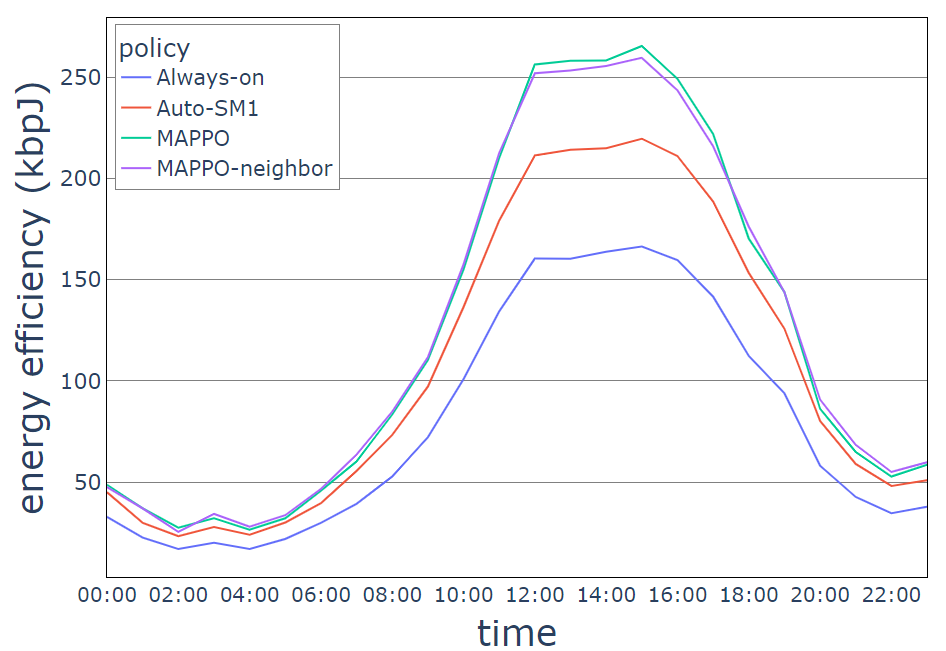}
    \vspace{-8mm}
    \caption{Comparison of energy efficiency. \label{fig:e-effi}}
\end{minipage}
\end{figure*}
When number of BSs is 19, we compare our proposed MAPPO-based algorithm and \emph{MAPPO-neighbor} algorithm with two baseline algorithms, i.e., the \emph{Always-on} algorithm and the \emph{Auto-SM1} algorithm. In the first algorithm, all BSs are kept active with all antennas turned on at all times. In the second algorithm, BSs without any associated UEs will be automatically switched to SM 1, and be promptly activated once a UE enters its signal coverage. Moreover, the antenna switching is the only action trained with MAPPO-based algorithm.

The temporal variations of the performance metrics are shown in Figs. 4-7.
In Fig. \ref{fig:data-rate}, the sum rate achieved by \emph{Auto-SM1} and \emph{Always-on} are nearly identical. This is due to the sleep mode of certain base stations, which, when reducing their transmission power, also decreases interference.
However, both MAPPO policies achieve a higher sum data rate compared to both baselines, which is attributed to the reduced interference received by users through the deactivation of an appropriate number of BSs and antennas. Sleeping also reduces PC, as demonstrated in Fig. \ref{fig:power}. It shows that the PC of the MAPPO policy is higher during the daytime and lower at night compared to \emph{Auto-SM1}, while \emph{MAPPO-neighbor} policy maintains nearly minimal power consumption throughout the day. 

In Fig. \ref{fig:drop-ratio}, we can observe that both MAPPO policies increase the drop ratio, since they strike a balance between QoS and PC, ensuring that the drop ratio remains at an acceptable level while sacrificing a small portion of the user experience in an effort to minimize energy consumption (deactivating BSs and antennas). Although \emph{MAPPO-neighbor} achieves a lower data rate than MAPPO during the daytime as shown in Fig. \ref{fig:data-rate}, Fig. \ref{fig:drop-ratio} shows that it performs better in terms of drop ratio. The reason for this is the \emph{MAPPO-neighbor} policy learns to enable more UEs' successful transmissions while reducing the energy allocated to existing UEs, as their actual data rates may exceed their requirements. This also explains why \emph{MAPPO-neighbor} consumes less power while has the same level of energy efficiency as MAPPO. 

In Fig. \ref{fig:e-effi}, we can observe that both MAPPO policies achieve better energy-saving performance compared to the \emph{ Always on} algorithm and the \emph{Auto-SM1} algorithm throughout the entire time period - their energy efficiencies are the highest, especially during high traffic hours. This is because the MAPPO policies have the ability to dynamically regulate the number of activated antennas and place the BSs in deeper SMs when necessary. By doing so, the policy optimizes the utilization of the limited available power for each BS, thereby maximizing the UEs' QoS by allocating more power for data transmission. 


\section{Conclusions}
In this paper, we have investigated the potential of MARL in the application of power consumption minimization of multiple massive MIMO BSs while preserving the overall user QoS by making decisions on the multi-level ASM and antenna switching of these BSs.
A DEC-POMDP has been formulated to characterize the dynamic network status. 
Then, a MAPPO-based algorithm has been proposed to obtain the joint sleep mode and antenna switching policy. 
Simulation results demonstrate that our proposed algorithm, \emph{MAPPO-neighbor} policy, can reduce the power consumption by approximately 8.7\% during the low-traffic hours, in comparison to the \emph{Auto-SM1} algorithm. Moreover, the energy efficiency can be improved mainly at high-traffic hours by approximately 19\%.

\section*{Acknowledgement}

This work was supported in part by the CELTIC-NEXT Projects, AI4Green and RAI6-Green, with funding received from Vinnova, Swedish Innovation Agency.

\bibliographystyle{IEEEtran}
\bibliography{references}

\end{document}